\documentclass[jcp,a4paper,superscriptaddress,floatfix,twocolumn]{revtex4}

\usepackage{graphics} 
\usepackage{latexsym} 
\usepackage{dcolumn}

\usepackage{amsfonts}
\usepackage{amsmath}
\usepackage{amsthm}

\begin{document} 
 
%\draft
 
\title{A density--functional study of interfacial properties of colloid--polymer mixtures\footnote{This paper is dedicated to David Chandler on the occasion of his 60th birthday.}} 
 
\author{A.\ Moncho-Jord\'{a}\footnote{Author to whom correspondence should be addressed. E-mail: moncho@ugr.es}} 
\affiliation{Departamento de F\'{\i}sica Aplicada, Universidad de Granada, Facultad de Ciencias, Campus Fuentenueva S/N, 18071 Granada, Spain.} 
 
\author{J.\ Dzubiella} 
 \affiliation{Department of Chemistry, University of Cambridge, Lensfield Road, Cambridge CB2 1EW, United Kingdom.}

\author{J.\ P.\ Hansen} 
 \affiliation{Department of Chemistry, University of Cambridge, Lensfield Road, Cambridge CB2 1EW, United Kingdom.} 
 
\author{A.\ A.\ Louis} 
 \affiliation{Department of Chemistry, University of Cambridge, Lensfield Road, Cambridge CB2 1EW, United Kingdom.}

%\date{\today} 

\begin{abstract}

Interfacial properties of colloid--polymer mixtures are examined
within an effective one--component representation, where the polymer
degrees of freedom are traced out, leaving a fluid of colloidal
particles interacting via polymer--induced depletion
forces. Restriction is made to zero, one and two--body effective
potentials, and a free energy functional is used which treats colloid
excluded volume correlations within Rosenfeld's Fundamental Measure
Theory, and depletion--induced attraction within first--order
perturbation theory. This functional allows a consistent treatment of
both ideal and interacting polymers. The theory is applied to surface
properties near a hard wall, to the depletion interaction between two
walls, and to the fluid--fluid interface of demixed colloid--polymer
mixtures. The results of the present theory compare well with
predictions of a fully two--component representation of mixtures of
colloids and ideal polymers (the Asakura--Oosawa model), and allow a
systematic investigation of the effects of polymer--polymer
interactions on interfacial properties. In particular, the wall
surface tension is found to be significantly larger for interacting
than for ideal polymers, while the opposite trend is predicted for the
fluid--fluid interfacial tension.

\end{abstract} 
 
%\pacs{02.70.Lq, 61.43.Hv, 82.70.Dd, 05.40.+j} 
 
\maketitle 

\section{Introduction} 

Mixtures of colloidal particles and non-adsorbing polymers dispersed in a solvent provide experimentalists and theoreticians with a very flexible model system to explore the statics and dynamics of phase transitions, including fluid--fluid demixing, crystalization, gelation or glass transitions\cite{Poon02}, as well as the interfacial properties associated with phase coexistence. The flexibility of this model system stems from the fact that its properties can be easily tuned by varying, among others, the size ratio of the two components (e.g.\ by controlling the degree of polymerization), their concentrations, or the quality of the solvent which determines whether the polymer coils behave essentially like interacting self--avoiding walks (SAW), or more like ideal non--interacting polymers (under $\theta$--solvent conditions).

More specifically if, as will be done in this paper, one adopts an effective one--component representation (by tracing out the polymer degrees of freedom) the resulting effective interactions between the colloidal particles, obtained from the well--known depletion mechanism\cite{Liko01}, are easily tuned by varying the above physical parameters: the depth of the effective attraction between colloids is essentially controlled by the polymer concentration, while the range depends on the polymer size (radius of gyration). The resulting phase diagrams are very sensitive to changes in the depletion--induced pair potential between colloids\cite{Bolh02a}. Recent experimental\cite{Aart03} and theoretical\cite{Brad03,Aart04a} efforts have focused on interfacial and wetting properties of colloid-polymer mixtures, either near solid substrates (hard walls) or at fluid--fluid phase coexistence. The density profiles near a glass wall and the surface tension at fluid--fluid phase coexistence were measured\cite{Aart03} and very recently the first direct observation of capillary fluctuations at the fluid--fluid interface was reported\cite{Aart04}. Various versions of density functional theory (DFT) of inhomogeneous fluids have been used to determine the density profiles, adsorption and surface tension of colloid--polymer mixtures near a hard wall, either in the one--component\cite{Brad01} or two--component\cite{Brad02,Wess04} representations, or at the phase--separated fluid--fluid interface\cite{Brad00,Monc03}. Most of the theoretical work so far focused on mixtures of hard sphere colloids and ideal (non--interacting) polymers within the classic model of Asakura and Oosawa\cite{Asak58} and of Vrij\cite{Vrij76} (the AO model), although some attempts have been made to extend the DFT calculations to the case of non--ideal (interacting) polymer coils\cite{Wess04,Monc03}, pointing to very significant differences between the two situations.

This paper presents a unified DFT description of wall--fluid and fluid--fluid interfaces within an effective one--component representation. The DFT theory is a perturbative one with respect to the polymer--induced depletion attraction between hard sphere colloids, and applies to mixtures of colloids and non--interacting, as well as interacting polymer coils. Apart from a systematic comparison of the density profiles and surface tensions obtained for these two cases, the present theory also leads to the first estimate of the depletion potential between two walls induced by colloid-polymer mixtures.

The paper is organized as follows. Section~\ref{eff} briefly summarizes the effective one and two--component representations of colloid-polymer mixtures. The DFT formulation used in this paper is presented in section~\ref{dff}. Results for colloid density profiles near a hard wall in the presence of ideal or interacting polymers are presented in section~\ref{dphw}, while the resulting wall surface tension of the mixtures is calculated in section~\ref{wst}. The depletion potential between hard walls, induced by colloid--polymer mixtures is described in section~\ref{dihw}. The fluid--fluid interfacial properties calculated within the same DFT approximation are presented in section~\ref{ff}, while conclusions are drawn in section~\ref{conc}. 

\section{Effective one and two--component representations}
\label{eff}

Consider a binary mixture of $N_c$ colloidal particles and $N_p$
polymer coils in an external field (e.g.\ the confining field of a hard
wall). The total Hamiltonian of the system may be written as
\begin{equation}
H=H_{cc}+H_{pp}+H_{cp}+\Phi_c+\Phi_p
\end{equation}
where the colloid--colloid, polymer--polymer and colloid--polymer
terms $H_{cc}$, $H_{pp}$ and $H_{cp}$ will be assumed to be sums of pair
interactions between the centres of mass (CM) $\{ \vec{R}_i \}$ ($1
\leq i \leq N_c$) and $\{ \vec{r}_j \}$ ($1 \leq j \leq N_p$) of the
colloids and polymers, respectively. Within this assumption, the
individual monomer degrees of freedom of the polymer coils have been
traced out, and the resulting effective pair potential between polymer
CM's is a state--dependent free energy\cite{Bolh01}; similarly the
colloid--polymer pair potential is a state--dependent free energy
resulting from a statistical average over monomer degrees of freedom
for a fixed distance between the CM's of the two
particles\cite{Bolh02}. Hence,
\begin{eqnarray}
H_{cc} &=& \sum_{i<j}^{N_c}v_{cc}(R_{ij}) \nonumber \\
H_{pp} &=& \sum_{i<j}^{N_p}v_{pp}(r_{ij})  \\
H_{cp} &=& \sum_{i}^{N_c}\sum_{j}^{N_p}v_{cp}(|\vec{R}_i-\vec{r}_j|) \nonumber
\end{eqnarray}
while
\begin{eqnarray}
\Phi_c=\sum_i^{N_c}\varphi_c^0(\vec{R}_i) \ \ ; \ \ \Phi_p=\sum_i^{N_p}\varphi_p^0(\vec{r}_i)
\end{eqnarray}
where $\varphi_c^0$ and $\varphi_p^0$ are direct external interactions acting on
colloidal particles and polymers, respectively. Note that
$\varphi_p^0(\vec{r})$ is once more an effective potential acting on the
polymer CM.

The colloid--colloid pair potential will henceforth always be
considered to be of the hard sphere form. The AO model for
non--interacting polymers further assumes that
\begin{equation}
\label{nonad}
v_{\alpha \beta}(r) = \left\{\begin{array}{l@{\quad}l}
\infty &  \;\; r < \sigma_{\alpha \beta}   \\
0 &  \;\; r > \sigma_{\alpha \beta} \ ; \ \alpha,\beta =c \ \mbox{or} \ p 
\end{array} \right.
\end{equation}
with $\sigma_{cc}=2R_c$ ($R_c$ being the colloidal radius),
$\sigma_{pp}=0$ and $\sigma_{cp}=R_c+R_p$, where $R_p$ is the radius
of gyration of the polymer coils. A generalization of
Rosenfeld's Fundamental Measure Theory (FMT) free energy functional
for additive hard spheres\cite{Rose89} has been worked out for the
non--additive hard sphere mixture given by eq~\ref{nonad}
\cite{Schm00}, and applied to colloid--polymer interfacial
properties\cite{Brad03,Brad02}.

An alternative approach is to trace out the polymer degrees of freedom to derive an effective
Hamiltonian involving only the colloid degrees of freedom\cite{Liko01,Loui02c}. 
Within the semi--grand canonical ensemble, with fixed $N_c$ and fixed
chemical potential $\mu_p$ of the polymers (or equivalently, fixed
number density $\rho_p^r$ of the polymers in a reservoir), the
effective Hamiltonian is:
\begin{equation}
\label{effh}
H_c^{eff}=H_{cc}+\Phi_c+\Omega,
\end{equation}
where $\Omega$ is the grand potential of the inhomogeneous fluid of
polymers, which depends parametrically on the positions $\{\vec{R}_i
\}$ of the colloidal particles. $\Omega$ can be systematically be
expanded in terms corresponding to the number of  colloids\cite{Aart04,Brad01}:
\begin{equation}
\label{nterms}
\Omega=\sum_{n=0}^{N_c}\Omega_n.
\end{equation}
$\Omega_0$ is the free energy of the pure polymer solution at chemical
potential $\mu_p$,  in the volume $V$ of interest, $\Omega_1$ is the
free energy cost of inserting independent single colloids, $\Omega_2$
takes pairwise colloid-colloid interactions into account, and so forth
for higher order terms.  Such an expansion can also be carried out near a 
single wall of area $A$, where the first few terms are:
\begin{eqnarray}
\label{Omega012}
 \Omega_0 &=& -P_p(\mu_p) V + \gamma_{w,p}(\mu_p) A \nonumber \\
 \Omega_1 &=& N_c \omega_1(\mu_p) +\sum_{i=1}^{N_c}\varphi_c^{eff}(z_i) \nonumber \\
\beta \Omega_2 &=& \sum_{i<j}^{N_c}\beta v_{eff}(R_{ij}) 
\end{eqnarray}
where $\varphi_c^{eff}(z)$ is the effective wall--colloid depletion potential induced by the presence of polymers, $P_p(\mu_p)$ is the osmotic pressure of the bulk polymer solution, and
 $\gamma_{w,p}(\mu_p)$ is the surface tension induced by the flat wall. Surface tensions are defined throughout relative to the position of the hard wall, at $z=0$. Inserting a single colloid into a bulk polymer solution costs a
 free energy $\omega_1(\mu_p) = \Omega^{bulk}_1/N_c$, and $\varphi_c^{eff}(z)$ describes the correction
 to that insertion free energy when the colloid is at a distance $z$ from
 the wall.  $v_{eff}(R_{ij})$ is the effective interaction between
 two colloids, induced by the polymer solution.  The next higher order
 contributions to $\Omega$ include a a three--body colloid term, a
 three--body colloid-colloid-wall term, etc...  These will be ignored
 in the present work.

Brader et al.\ systematically worked out this
expansion for an AO mixture near a flat wall, finding~\cite{Brad01}: $\beta \Omega_0=
-\rho_p^r (V -R_pA)$ (since $\beta P_p(\mu_p) = \rho_p^r$ and $\beta \gamma_{w,p}^{id} =
\rho_p^r R_p$ for AO particles\cite{Loui02b}), $\beta \Omega_1 =
N_c\rho_p^r \frac{4}{3}\pi(R_c+R_p)^3 +\sum_{i=1}^{N_c}\beta
\varphi_{AO}(z_i)$, and $\beta \Omega_2 = \sum_{i<j}\beta v_{AO}(R_{ij})$.
$\beta \varphi_{AO}(z_i)$ 
acts on each colloid independently and takes the form:
\begin{equation}
\label{varphiAO}
\beta \varphi_{AO}(s)=-\frac{\eta_p^r}{4q^3}(2q-s)^2(3+q+s)\theta(s-2q),
\end{equation}
where $\eta_p^r=4\pi \rho_p^r R_p^3/3$ is the polymer packing fraction
in the reservoir,
 $q=R_p/R_c$ is the size ratio, and $\theta(x)$ is
the Heaviside step function ($\theta(x)=1$, $x<0$; $\theta(x)=0$,
$x>0$). $v_{AO}(R_{ij})$ is the Asakura--Oosawa depletion\cite{Asak58} potential between
two spheres in a bath of ideal polymers; defining now
$s=(R-2R_c)/R_c$, $v_{AO}$ is given by:
\begin{equation}
\label{vAO}
\beta v_{AO}(s)=-\frac{\eta_p^r}{16q^3}(2q-s)^2(6+4q+s)\theta(s-2q)
\end{equation}

The effective Hamiltonian for the AO model, with $\Omega$ restricted
to the 0, 1 and 2--body terms in eq~\ref{nterms} is strictly valid
only for size ratios $q<0.1547$; for larger size ratios, three and
more--body effective interactions between colloidal particles come
into play \cite{Brad01}, but explicit calculations of bulk properties
show that they do not play a significant role for $q$ up to $\approx
1$ \cite{Dijk99,Rote04}. Subsequent calculations will neglect more
than two--body interactions. For the AO model, $\Omega_0$ and the bulk
part of $\Omega_1$, $\Omega_1^{bulk} = N_c \omega_1$, do not affect the interfacial profiles, but do make
contributions to the surface tension $\gamma$ as will be shown in
subsequent sections.

The AO pair potential (eq~\ref{vAO}) assumes ideal polymers to be
spherical, so that the width of the depletion layer is always equal to
$2R_p$. However, polymers can \symbol{92}wrap" around colloidal
spheres, the more so the smaller the radius $R_c$ of the
latter. Hence, one expects the width of the depletion layer around a
colloid to shrink as $R_c$ decreases. This effect can be accounted for
by following the prescription of refs~\cite{Loui02,Meij94}, whereby the AO form (eq~\ref{vAO}) of the
effective pair potential holds, but  with a renormalized size ratio
$q^{\prime}$ and a renormalized polymer packing fraction
${\eta_p^r}^{\prime}$:
\begin{eqnarray}
\label{qetap}
q^{\prime} &=& \left( 1+\frac{6}{\sqrt{\pi}}q+3q^2 \right)^{1/3}-1 \nonumber \\
{\eta_p^{r}}^{\prime} &=&\left( \frac{q^{\prime}}{q} \right)^3\eta_p^r
\end{eqnarray}
 
For interacting polymers we can also carry out an expansion of
$\Omega$. Accurate expressions for $P_p$, $\gamma_{w,p}$, $\omega_1$
and $v_{eff}(r)$ are known and have been validated by computer
simulations for self--avoiding walk (SAW) chains\cite{Loui02,Loui02b},
so that $\Omega_0$, $\Omega_1$ and $\Omega_2$ directly follow from
eq~(\ref{Omega012}).  The Derjaguin approximation, which is reasonably
accurate for an ideal polymer depletant, is expected to remain valid
in the case of interacting polymers, at least for sufficiently small
$q$, so that the required wall--colloid effective depletion potential
$\varphi_c^{eff}(z_i)$ is given by
\begin{equation}
\label{varphis}
\beta \varphi_{int}(s) \approx 2\beta v_{int}(s),
\end{equation}
where the effective potential between two colloids induced by 
interacting polymers is well approximated by the semi-empirical form
\begin{equation} 
\label{Vs} 
\beta v_{int}(s)=-\pi \frac{\beta R_c^2 \gamma_{w,p}(\rho_p^r)}{d(\rho_p^r)}\left[s-d(\rho_p^r) \right]^2 \theta(s-d)
\end{equation} 
Here $d=D/R_c$ and $D$ is the range of the depletion potential:
\begin{equation} 
D(\rho_p^r)=\sqrt{\pi}\frac{\gamma_{w,p}(\rho_p^r)}{P_p(\rho_p^r)}.
\frac{q^{\prime}}{q}
\end{equation} 
  Because $v_{int}(r)$
depends on polymer density only through $\rho_p^r$ (i.e.\ $\mu_p$),
its form does not change when the polymer density is inhomogeneous.
In contrast to the AO model, higher order colloid--colloid and
colloid--wall terms are relevant for any size ratio.

The effective one--component representation of colloid--polymer
mixtures is thus fully defined, both for ideal and interacting
polymers. The next step is to define the density functional appropriate
for the description of the inhomogeneous effective one--component
model.

\section{Density functional formulation}
\label{dff}

Given the effective Hamiltonian specified by eqs~\ref{effh} and \ref{Omega012}, one may construct an approximate free energy density functional to investigate the properties of the inhomogeneous effective one--component system of colloidal particles. The latter interacts via a hard sphere repulsions (for $r<2R_c$) and a polymer--induced depletion attraction that will be described within first order perturbation theory\cite{Tang91}, which is expected to be accurate for not too small values of $q$ (i.e.\ for sufficiently long--range attractions). The intrinsic free--energy functional is then conveniently split into ideal, hard sphere and perturbation parts; for the hard core part we adopt the very accurate \symbol{92}White Bear" version\cite{Roth02} of Rosenfeld's \symbol{92}Fundamental Measure" (FMT) functional\cite{Rose89}
\begin{equation}
\label{FPDFT}
F \left[ \rho_c(\vec{r})\right]= F_{id} \left[ \rho_c(\vec{r})\right]+F_{FMT} \left[ \rho_c(\vec{r})\right]+F_1 \left[ \rho_c(\vec{r})\right]
\end{equation}
The ideal contribution is:
\begin{equation}
F_{id} \left[  \rho_c \right]=k_B T \int d\vec{r}\rho_c(\vec{r})\left[ \ln(\Lambda_c^3\rho(\vec{r}))-1 \right]
\end{equation}
where $\Lambda_c$ is an irrelevant colloidal length scale. The FMT--hard core contribution is of the \symbol{92}weighted density" type, namely
\begin{eqnarray}
\label{FFMT}
\beta F_{FMT} \left[  \rho_c \right] &=&\int d\vec{r} [ \Phi_1\left( \{ n_j(\vec{r})\} \right)+\Phi_2\left( \{n_j(\vec{r})\} \right) \nonumber \\
&+& \Phi_3\left( \{n_j(\vec{r})\} \right) ]
\end{eqnarray}
where the $n_j(\vec{r})$ are weighted densities of the form
\begin{equation}
\label{nj}
n_j(\vec{r})=\int \omega^{(j)}(\vec{r}-\vec{r}^{\ \prime})\rho_c(\vec{r}^{\ \prime})d\vec{r}^{\ \prime}
\end{equation}
The functions $\Phi_i$ and the weight functions $w^{(j)}(\vec{r})$ are defined in Appendix~A. Finally, the first order perturbation term reads:
\begin{equation}
\label{F1}
F_1 \left[  \rho_c \right]=\int d\vec{r} \rho_c(\vec{r})\Psi_{1}(\vec{r})
\end{equation}
with
\begin{equation}
\label{Psi1}
\Psi_{1}(\vec{r})= \frac{1}{2}\int d\vec{r}^{\ \prime}\rho_c(\vec{r}^{\ \prime})v(| r-r^{\ \prime} |)g_{hs}(| r-r^{\ \prime} |,\bar{\rho}_c(\vec{r},\vec{r}^{\ \prime})) 
\end{equation}
where $v(r)=v_{AO}(r)$ or $v_{int}(r)$ is the depletion potential induced by ideal ($AO$) or interacting ($int$) polymer coils; $g_{HS}$ is the pair distribution function of the homogeneous reference hard sphere fluid, evaluated at an intermediate density between the two points $\vec{r}$ and $\vec{r}^{\ \prime}$: 
\begin{equation}
\label{rhocbar}
\bar{\rho}_c=\frac{\bar{\rho}_{\nu}(\vec{r})+\bar{\rho}_{\nu}(\vec{r}^{\ \prime})}{2}
\end{equation}
with
\begin{equation}
\label{rhocbarnu}
\bar{\rho}_{\nu}(\vec{r})=\frac{3}{4 \pi R_{\nu}^3}\int_{| r-r^{\ \prime} |<R_{\nu}}\rho_c(\vec{r}^{\ \prime})d\vec{r}^{\ \prime}.
\end{equation}
$\bar{\rho}_{\nu}(\vec{r})$ is a smoothed density profile around $\vec{r}$; the radius $R_{\nu}$ is of the order of $R_c$, and results are not expected to be very sensitive to the precise value of $R_{\nu}$. Following the earlier experience of ref~\cite{Soko92} we have chosen $R_{\nu}=1.6R_c$. The form \ref{Psi1} is the generalisation of the standard first order thermodynamic theory \cite{Hans86} to inhomogeneous fluids. The choices given in eq \ref{rhocbar} and \ref{rhocbarnu} have proved very adequate in many DFT calculations of fluids near hard walls.

In the homogeneous limit, where the density profile reduces to the bulk density, the free energy given by eq~\ref{FPDFT} goes over to the Helmholtz free energy of the fluid phase calculated within first order thermodynamic perturbation theory\cite{Chan83}, which leads to reasonable phase diagrams of colloid--polymer mixtures\cite{Rote04}. The familiar generalized van der Waals mean field approximation amounts to setting $g_{HS}(r)=\theta (2R_c-r)$, which leads of course to a considerable simplification of the DFT calculations. 

Keeping in mind that the total effective Hamiltonian of the colloids is given by eqs~\ref{effh}--\ref{Omega012}, with $\Omega = \Omega_0+\Omega_1+\Omega_2$, the grand potential functional to be minimized with respect to the density profile $\rho_c(\vec{r})$ is the sum of the intrinsic free energy functional \ref{FPDFT} and of the contributions from the external fields:
\begin{eqnarray}
\label{omegagrand}
\Omega_{\varphi}\left[ \rho_c \right] &=& F\left[ \rho_c \right]+\int \rho_c(\vec{r})\left[\varphi_c(\vec{r})-\mu_c^0 \right]d\vec{r} \nonumber \\
&+&\Omega_0+\Omega_1^{bulk}
\end{eqnarray}
where $\mu_c^0$ is the bulk chemical potential of the colloids, 
while $\Omega_0$ and $\Omega_1^{bulk} = N_c \omega_1$ are defined in
eq~\ref{Omega012}. Since the latter are constants, they have no
influence on the equilibrium profile $\rho_c(\vec{r})$, but they will
contribute to the equilibrium value of the grand potential, and hence
to the surface tension. $\varphi_c(\vec{r}) = \varphi_c^0(\vec{r})
+\varphi_c^{eff}(\vec{r})$ is the sum of the external potential acting
directly on the colloidal particles ($\varphi_c^0(\vec{r})
$) and of the polymer--induced
depletion contribution, $\varphi_c^{eff}(\vec{r})$, given by
eq~\ref{varphiAO} or \ref{varphis} for ideal and interacting polymers
respectively.

The Euler--Lagrange equation of the variational problem reads as:
\begin{equation}
\label{exprb}
\frac{\delta \Omega_\varphi[\rho_c(\vec{r})]}{\delta \rho_c(\vec{r})}=\frac{\delta F[\rho_c(\vec{r})]}{\delta \rho_c(\vec{r})} +\varphi_c(\vec{r})-\mu_c^0=0
\end{equation}
$F$ is the sum of ideal ($F_{id}$) and excess ($F_{ex}=F_{FMT}+F_1$) parts; defining the local excess chemical potential as 
\begin{equation}
\label{mucex}
\mu_c^{ex}(\vec{r})= \frac{\delta F_{ex}[\rho_c(\vec{r})]}{\delta \rho_c(\vec{r})}=\frac{\delta F_{FMT}[\rho_c(\vec{r})]}{\delta \rho_c(\vec{r})}+\frac{\delta F_{1}[\rho_c(\vec{r})]}{\delta \rho_c(\vec{r})},
\end{equation}
the Euler--Lagrange equation may be cast in the form:
\begin{equation}
k_BT\ln \left(\Lambda_c^3\rho_c(\vec{r})\right]+\mu_c^{ex}(\vec{r})=\mu_c^0-\varphi_c(\vec{r})
\end{equation}
so that the density profile satisfies the non--linear equation:
\begin{equation}
\label{rhorho0}
\rho_c(\vec{r})=\rho_c^{0}\exp \{-\beta \left[\varphi_c(\vec{r})+\mu_c^{ex}(\vec{r}))- \mu_c^{0,ex}  \right] \}
\end{equation}
where $\mu_c^{0,ex}$ is the the excess part of the bulk chemical potential of the colloids. The coupled equations~\ref{mucex} (with $F_{FMT}$ and $F_1$ given by eqs~\ref{FFMT} and \ref{F1})  and \ref{rhorho0} must be solved numerically by standard iterative procedures.

Since only planar interfaces will be considered, the external potential, the density profile and local chemical potential all depend only on $z$, the coordinate orthogonal to the plane of the interface. The corresponding simplified expression for the local excess chemical potential $\mu_c^{ex}$ \cite{Wade00} is given in Appendix~B, together with the weight functions $w^{(j)}(z-z^{\prime})$ appropriate for the one--dimensional problem.

Once the equilibrium profile $\rho_c(z)$ has been determined, it can be substituted into the expression of the functional \ref{omegagrand} to determine the equilibrium grand potential $\Omega$. Subtraction of the bulk contribution then allows the surface tension $\gamma$ to be calculated, according to the definition:
\begin{equation}
\label{gammaP}
\gamma=\frac{\Omega+PV}{A}
\end{equation}
where $P$ is the total bulk osmotic pressure of the mixture and $A$ the total area of the planar interface.

\section{Density profiles near a hard wall}
\label{dphw}

\begin{figure} 
\center\resizebox{0.47\textwidth}{!}{\includegraphics{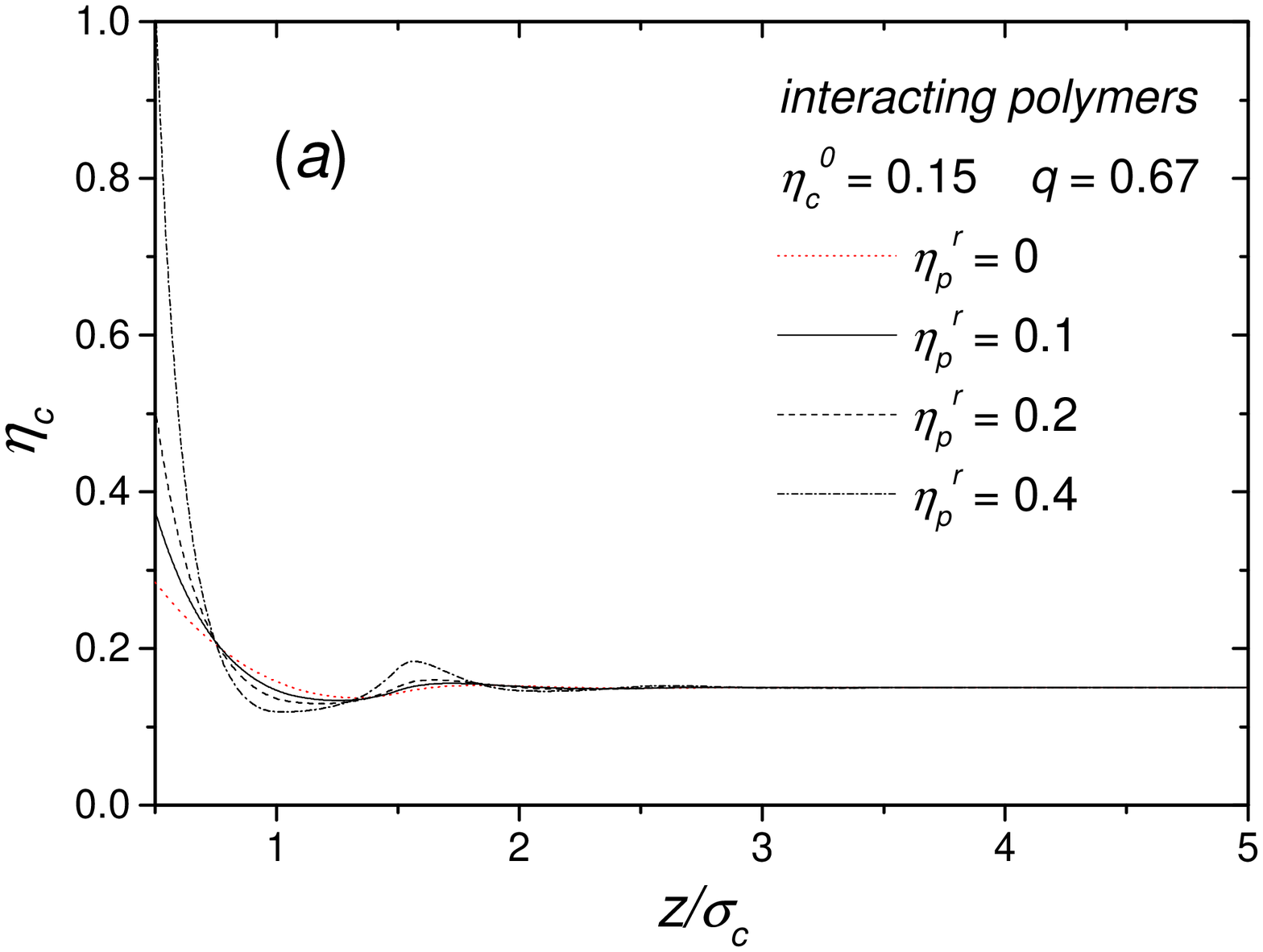}}
\center\resizebox{0.47\textwidth}{!}{\includegraphics{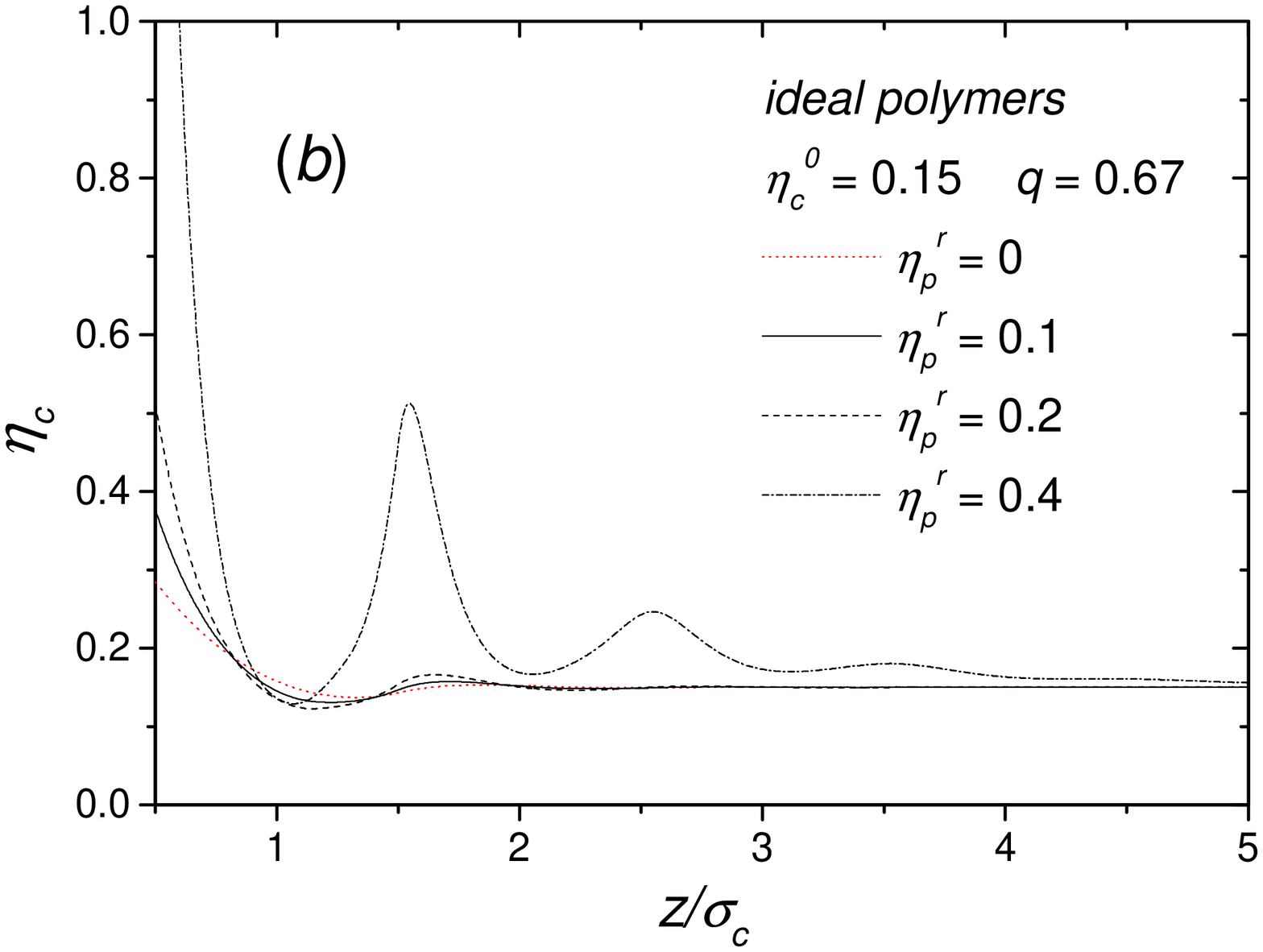}} 
\caption{\label{RHO1} Colloid density as a function of the distance to the hard wall for mixtures of colloids and (a) interacting polymers and (b) ideal polymers, at different polymer concentrations. In both figures the bulk colloid packing fraction and the size ratio were set at $\eta_c^{0}=0.15$ and $q=0.67$.} 
\end{figure}

Consider first the case of a colloid--polymer mixture near a hard wall placed at $z=0$. The density profiles were calculated by solving the one--dimensional versions of eqs~\ref{mucex} and \ref{rhorho0} (see Appendix~B). A systematic comparison is made between profiles for ideal and interacting polymers, over a range of colloid and polymer packing fractions $\eta_c^0$ and $\eta_p^r$ and for size ratios $q=0.34$, 0.5, 0.67, 0.85 and 1.05. Figures~\ref{RHO1}a,b compare the density profiles calculated for a fixed colloidal density and several polymer concentrations, for interacting and ideal polymers. As expected, the colloid adsorption at contact increases dramatically with increasing polymer concentration, due to the enhanced effective depletion--induced attraction to the wall. The effect is stronger for ideal polymers, which also lead to strong layering at the highest packing fraction hinting at a possible layering transition. The enhanced adsorption and layering are easily understood, since the polymer induced attraction is substantially stronger for ideal polymers (cf. eq~\ref{varphiAO}) compared to interacting polymers (eqs \ref{Vs} and \ref{varphis}), at the same $R_p$, $q$ and $\rho_p^r$\cite{Loui02b}. The structured profile at $\eta_p^r=0.4$ in Figure~\ref{RHO1}b is for a thermodynamic state close to the fluid--fluid binodal (on the colloid--poor \symbol{92}vapour side") calculated from the same free energy model (see ref~\cite{Rote04}) giving further support to the possibility of a layering transition\cite{Evans01,Dijk02}.

\begin{figure} 
\center\resizebox{0.47\textwidth}{!}{\includegraphics{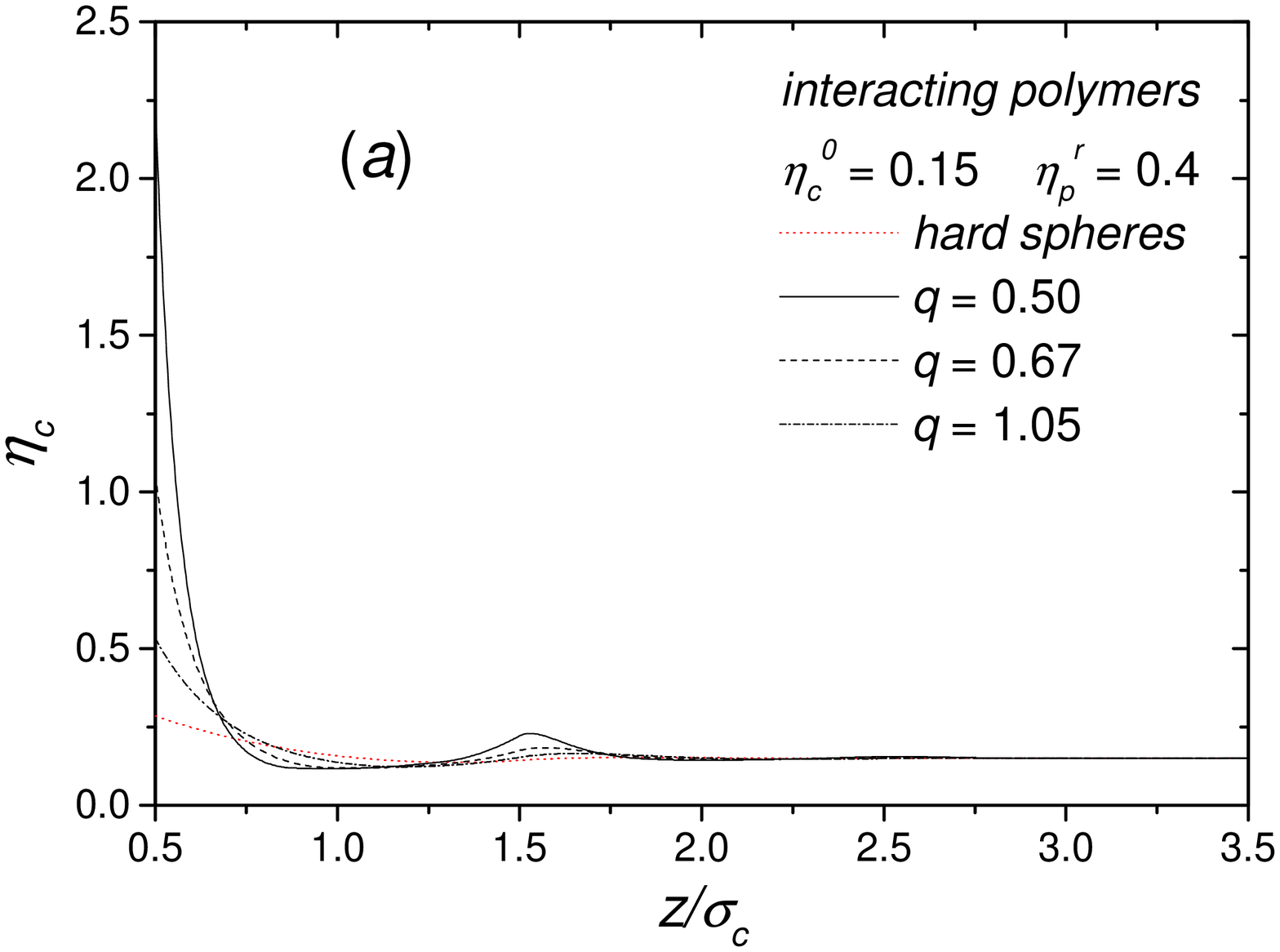}}
\center\resizebox{0.47\textwidth}{!}{\includegraphics{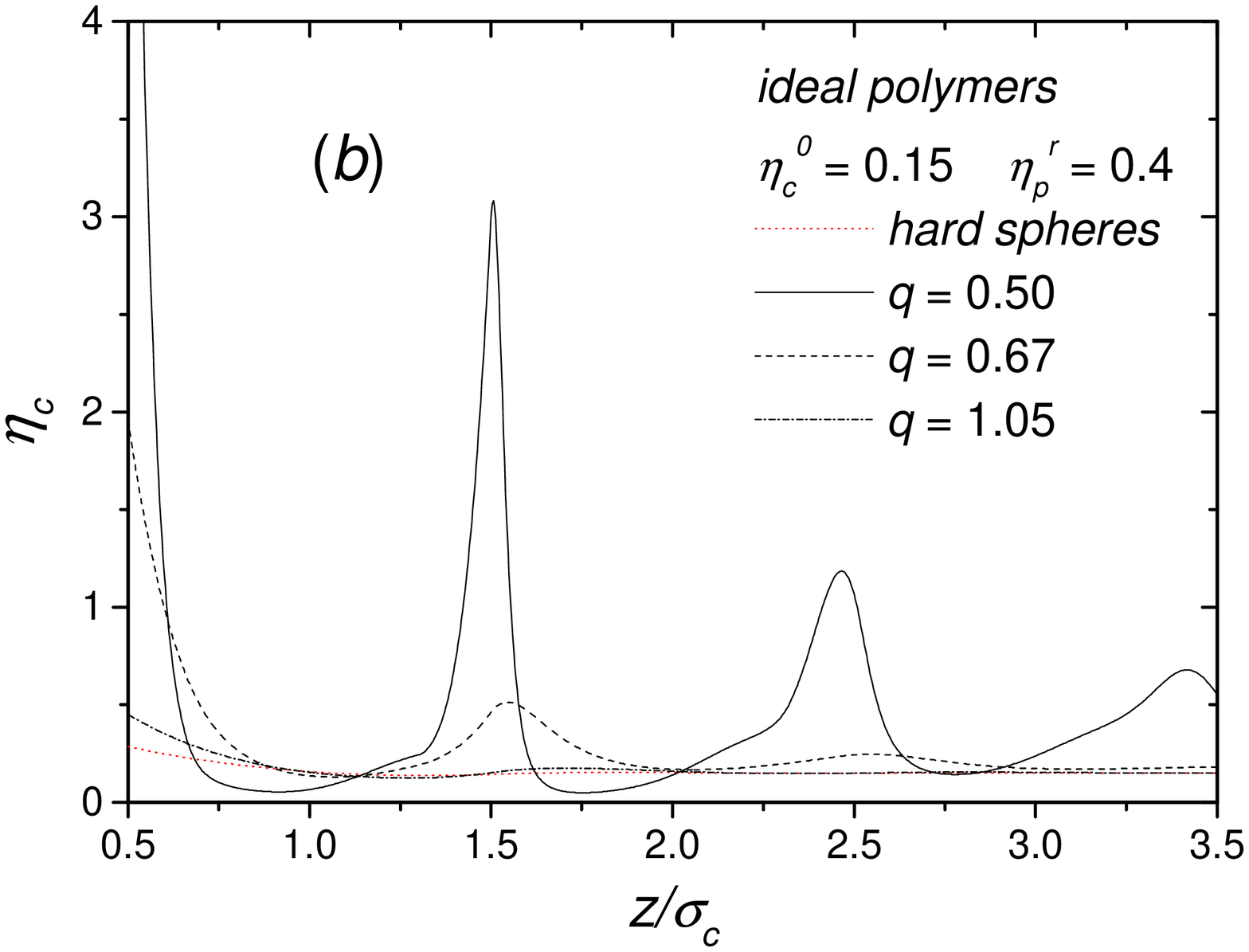}} 
\caption{\label{RHO2} Effect of the polymer--colloid size ratio, $q$, on the density profile near a hard wall for $\eta_c^{0}=0.15$ and $\eta_p^r=0.4$ induced by (a) interacting and (b) ideal polymers. In both pictures, the density profile of pure one--component hard spheres (without depletant) at $\eta_c^{0}=0.15$ as been included for comparison.} 
\end{figure}

The effect of the range of the depletion attraction is illustrated in
 Figures~\ref{RHO2}a,b for interacting and ideal polymers
 respectively. The trend of the density profiles with increasing $q$
 is opposite that observed with increasing $\eta_p^r$. The
 adsorption at contact is now strongest for the smallest size ratio.
 This is essentially due to the fact that the wall--colloid depletion
 potential at contact, $\varphi_c^{eff}(0)$, drops as the polymer density
 decreases. Since $\rho_p^r \sim \eta_p^r/q^3$, a small increase in
 the size ratio $q$ implies a big reduction in polymer density, and also 
in $\varphi^{eff}_c(0)$. Calculations carried out at higher colloid packing fractions show that the effect of polymer--induced wall--colloid attraction on the density profiles is much reduced as $\eta_c^0$ increases, due to the predominance of purely excluded volume effects under high density conditions\cite{Chan83,Hans86}

\section{Wall surface tension}
\label{wst}

Once the colloid density profiles have been calculated, the surface tension
may be determined by substituting $\rho_c(z)$ into
eq~\ref{omegagrand}, and then applying relation~\ref{gammaP}. Some
care has to be taken in treating the bulk zero and one--body
contribution $\Omega_0$ and $\Omega_1^{bulk}$. These also contribute
to the bulk properties (calculated by letting $\rho_c(z)$ go to the
bulk value $\rho_c^0$), and in particular to the colloid chemical
potential $\mu_c^0$ and to the osmotic pressure of the mixture. Thus,
for the chemical potential of the bulk one obtains:
\begin{eqnarray}
\label{mutot}
\mu_c^0=\mu_c({\rho_c^0})+\omega_1
\end{eqnarray}
where $\omega_1$ is already defined in eq \ref{Omega012}.

In a similar fashion, the total osmotic pressure is the sum of the colloid contribution $P_c(\rho_c^0,\rho_p^r)$ (calculated by differentiating the homogeneous limit of the intrinsic free energy \ref{FPDFT} with respect to $\rho_c^0$) and of the osmotic pressure of the polymer reservoir:
\begin{equation}
\label{Ptot}
P=P_c(\rho_c,\rho_p^r)+P_p(\rho_p^r)
\end{equation}

Hence,
\begin{eqnarray}
\label{newOmega}
\lefteqn{\frac{\Omega+PV}{A}=\frac{\Omega_0+\Omega_1+\Omega_2+PV}{A}=} \nonumber \\
&=& \{ \frac{-P_pV+\gamma_{w,p}A}{A} \} + \\
&+& \{ \omega_1(\rho_p^r)\int dz\rho_c(z) +\int dz \rho_c(z)\varphi_c(z) \} +\nonumber \\
&+& \{ \frac{F\left[ \rho_c(z) \right]}{A} -\mu_c^0\int dz \rho_c(z) \} +\frac{PV}{A} \nonumber
\end{eqnarray}
where $F\left[ \rho_c(z) \right]$ is the colloid contribution to the
 inhomogeneous free energy, given by eq~\ref{FPDFT}.  Substituting
 eqs~\ref{mutot} and \ref{Ptot} shows that the bulk contributions to
 $\Omega_0$ and $\Omega_1$ are exactly cancelled so that the final
 expression for the surface tension is given by:
\begin{eqnarray}
\label{gammaDFT}
\lefteqn{\gamma_w = \lim_{L\rightarrow \infty}\frac{\Omega +PV}{A}= \gamma_{w,p} +} \\
&+& \lim_{L\rightarrow \infty}\int_0^L dz\{ f\left[ \rho_c(z) \right] +\rho_c(z)[\varphi_c(z)-\mu_c(\rho_c^0)]+P_c\} \nonumber
\end{eqnarray}
where $f = F/V$ is the intrinsic colloid free energy density
 \ref{FPDFT}. In other  words, 
the surface tension may be written as the sum of a colloid
 contribution $\gamma_{w,c}$ (which would be obtained by ignoring
 $\Omega_0$ and $\Omega_1^{bulk}$ in eq~\ref{omegagrand}) and of the
 contribution of the polymers under reservoir conditions:
\begin{equation}
\label{gammatot}
\gamma_w=\gamma_{w,c}(\rho_c^0,\rho_p^r)+\gamma_{w,p}(\rho_p^r)
\end{equation}

\begin{figure} 
\center\resizebox{0.47\textwidth}{!}{\includegraphics{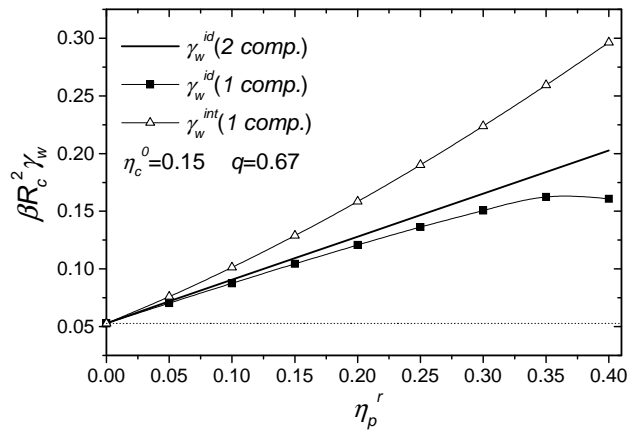}} 
\caption{\label{GAMMAW9} Wall surface tension for ideal (squares) and interacting polymers (triangles) in the one-component representation, and for ideal polymers in the two component representation (solid line), as a function of $\eta_p^r$ ($\eta_c^0=0.15$ and $q=0.67$). The dotted horizontal line is for hard sphere colloids only (without depletant).} 
\end{figure}

Equation~\ref{gammatot} is exact within an effective one--component
representation of a mixture of colloids and polymers where
$\Omega$ in eq~\ref{nterms} is truncated at $\Omega_2$. For
$\gamma_{w,c}(\rho_c^0,\rho_p^r)$ of interacting (SAW) polymers, we
use the results of ref~\cite{Loui02}. The resulting total surface
tensions for the cases of ideal and interacting polymers are plotted
as a function of the polymer reservoir packing fraction in
Figure~\ref{GAMMAW9}, and compared to the predictions of the
two--component representation (with ideal polymers), calculated within
scaled particle theory\cite{Wess04}, modified so as to account for the
\symbol{92}polymer wrapping" effect, according to the prescriptions of
Eq.~\ref{qetap}.

The agreement between the surface tensions calculated within the effective one and two--component representations is seen to be good for low densities and to deteriorate somewhat beyond. The surface tension with interacting polymers is substantially larger (up to a factor of two at the highest polymer concentrations) compared to the ideal polymer case. This is because the colloid--wall surface tension is higher with interacting polymers, due to the weaker effective wall--colloid attraction, and the wall--polymer surface tension is also higher for interacting polymers\cite{Loui02b}.

\begin{figure} 
\center\resizebox{0.47\textwidth}{!}{\includegraphics{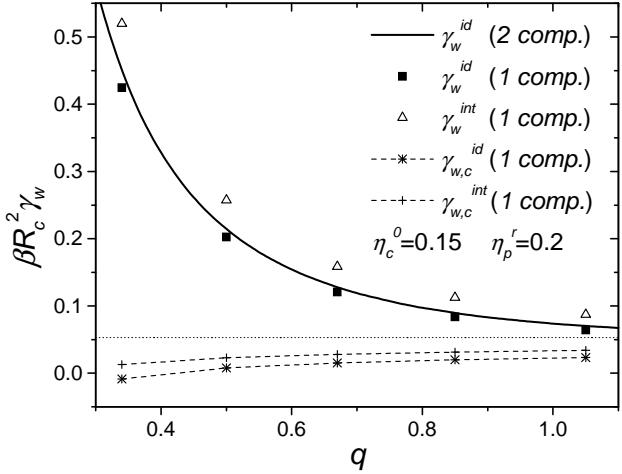}}
\caption{\label{GAMMAW10} Wall surface tension for ideal (squares) and
 interacting (triangles) polymers in the one-component representation,
 and for ideal polymers in the two--component representation (solid
 line), as a function of the size ratio, $q$ ($\eta_c^0=0.2$ and
 $\eta_p^r=0.2$). The dotted horizontal line is for pure hard
 spheres, and the colloidal contributions to the wall surface tension, $\gamma_{w,c}$, are also shown for comparison.}
\end{figure}

\begin{figure} 
\center\resizebox{0.47\textwidth}{!}{\includegraphics{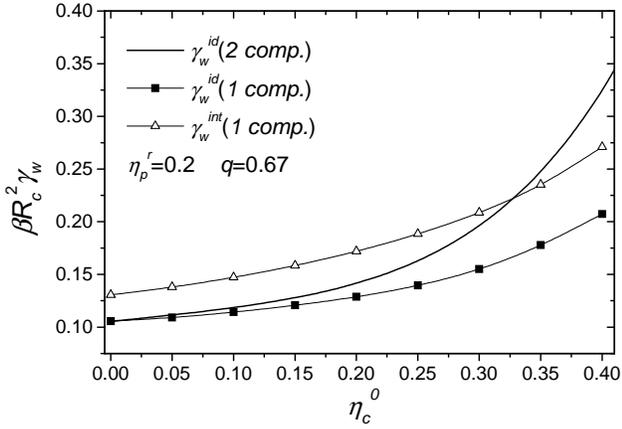}} 
\caption{\label{GAMMAW11} Wall surface tension for ideal (squares) and
interacting (triangles) polymers in the one-component representation,
and for ideal polymers in the two--component representation (solid
line), as a function of the bulk colloid packing fraction, $\eta_c^0$
($\eta_p^r=0.2$ and $q=0.67$).}
\end{figure}

Figure~\ref{GAMMAW10} illustrates the variation of the surface tension
 with size ratio $q$.  For fixed values of $\eta_c^0$ and $\eta_p^r$,
 the surface tension is seen to decrease rapidly as $q$ increases, and
 to tend to the result for hard spheres when $q>1$. Finally the
 variation of the surface tension with colloid packing fraction (for
 fixed $q$ and $\eta_p^r$) is shown in Figure~\ref{GAMMAW11}. The
 agreement between the effective one and two--component
 representations found at low--packing fractions is seen to
 deteriorate rapidly as $\eta_c^0$ increases, thus illustrating the
 break--down of the effective one--component model for highly
 concentrated colloidal suspensions.


\section{Depletion interaction between hard walls}
\label{dihw}

The depletion potential per unit area induced by a colloid--polymer
mixture between two hard walls of area $A$ separated by $L$ is
determined by:
\begin{equation}
W(L)=\frac{\Omega (L)-\Omega (\infty)}{A}
\end{equation}

Clearly, $W(L)\rightarrow 0$ as $L\rightarrow \infty$, while $W(L=0)=
-2\gamma_w$, because when the walls come into contact two fluid--wall
interfaces are destroyed. For arbitrary $L$, the colloid contribution
$W_c$ to $W(L)$ obtained by momentarily ignoring the bulk
contributions $\Omega_0$ and $\Omega_1^{bulk}$ to the total grand
potential, reads, by straightforward analogy with
expression~\ref{gammaDFT} for the surface tension:
\begin{eqnarray}
\label{WcDFT}
\lefteqn{W_c(L) = \int_0^L dz\{ F\left[ \rho_c(z) \right] } \nonumber \\
\ \ \ \ \ &+& \rho_c(z)[\varphi_c(z)-\mu_c(\rho_c^0)]+P_c\} -2\gamma_{w,c}
\end{eqnarray}

\begin{figure} 
\center\resizebox{0.47\textwidth}{!}{\includegraphics{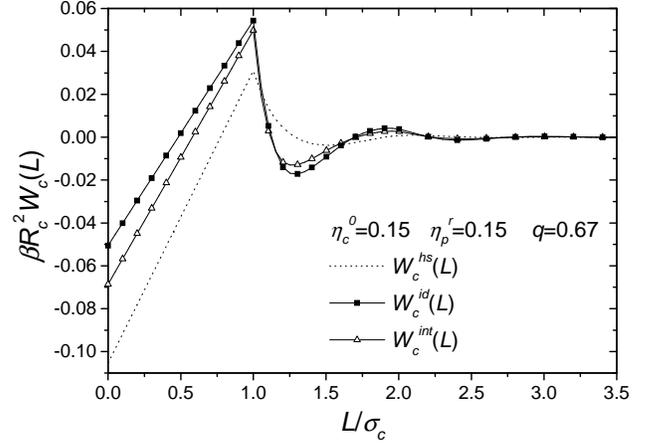}} 
\caption{\label{WL1} Effective depletion potential between two flat parallel walls as a function of the wall-wall distance, $L$, induced by pure hard colloids (dotted line), colloids plus ideal polymers (triangles) and colloids plus ideal polymers (squares). $\eta_c^0=0.15$, $\eta_p^r=0.15$ and $q=0.67$.} 
\end{figure}

Figure~\ref{WL1} shows the comparison of the depletion potentials
$W_c(L)$ calculated for a pure hard sphere depletant (i.e.\
$\eta_p^r=0$) and for hard sphere colloids and ideal or interacting
polymers, within the present effective one--component
representation. The extra attraction between the walls and colloidal
particles forces the latter into the region between the two plates,
enhancing the potential barrier of the depletion potential at
$L=R_c$. The effect is stronger for ideal than for interacting
polymers, because effective wall--colloid and colloid--colloid
attractive interactions are enhanced in going from interacting to
ideal polymers. For $L<2R_c$ the hard colloids are excluded from the
space between the walls, and the depletion potential is entirely
controlled by the external, unbalanced osmotic pressure which pushes
the walls together, i.e.:
\begin{equation}
W_c(L < 2R_c)=P_cL-2\gamma_{w,c}
\end{equation}

\begin{figure} 
\center\resizebox{0.47\textwidth}{!}{\includegraphics{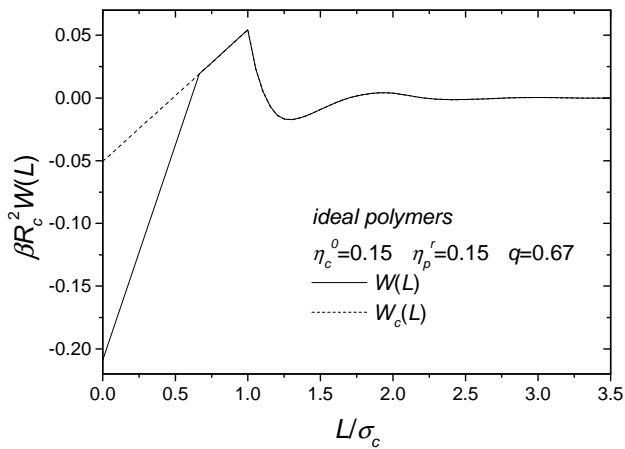}} 
\caption{\label{WL2} Comparison between the wall-wall depletion potential in a mixture of colloids and ideal polymers, with and without the bulk contributions to the free energy ($W$ and $W_c$, respectively). $\eta_c^0=0.15$, $\eta_p^r=0.15$ and $q=0.67$.} 
\end{figure}

To obtain the total depletion potential, the  volume terms $\Omega_0$
and $\Omega_1^{bulk}$ must be included  in the grand potential.
It is easily verified that, within the truncation of $\Omega$ at $\Omega_2$, 
i.e.\ ignoring colloid--colloid--polymer and similar higher order terms,
the depletion potential takes the form:
\begin{equation}
\label{WLtotc}
W(L) = \left\{\begin{array}{l@{\quad}l}
W_c(L)+P_p^rL-2\gamma_{w,p}(\rho_p^r) &  \;\; L < D_w^p  \\
W_c(L) &  \;\; L \geq D_w^p   
\end{array} \right.
\end{equation}
where $P_p^r=k_BT\rho_p^r$ is the osmotic pressure of the polymers in
the reservoir, $\gamma_{w,p}=k_BT\rho_p^rR_p$ is the corresponding
wall--polymer surface tension, and $D_w^p =  2
\gamma_{w,p}/P_p$ is the range of the polymer
depletion potential, which takes the value $D_w^{id} = 2 R_p$ for AO particles.
 The resulting total depletion potential for ideal polymers is shown in 
Figure~\ref{WL2}. It remains continuous at $L=2R_p$, but the resulting
force is discontinuous at that separation.

\begin{figure} 
\center\resizebox{0.47\textwidth}{!}{\includegraphics{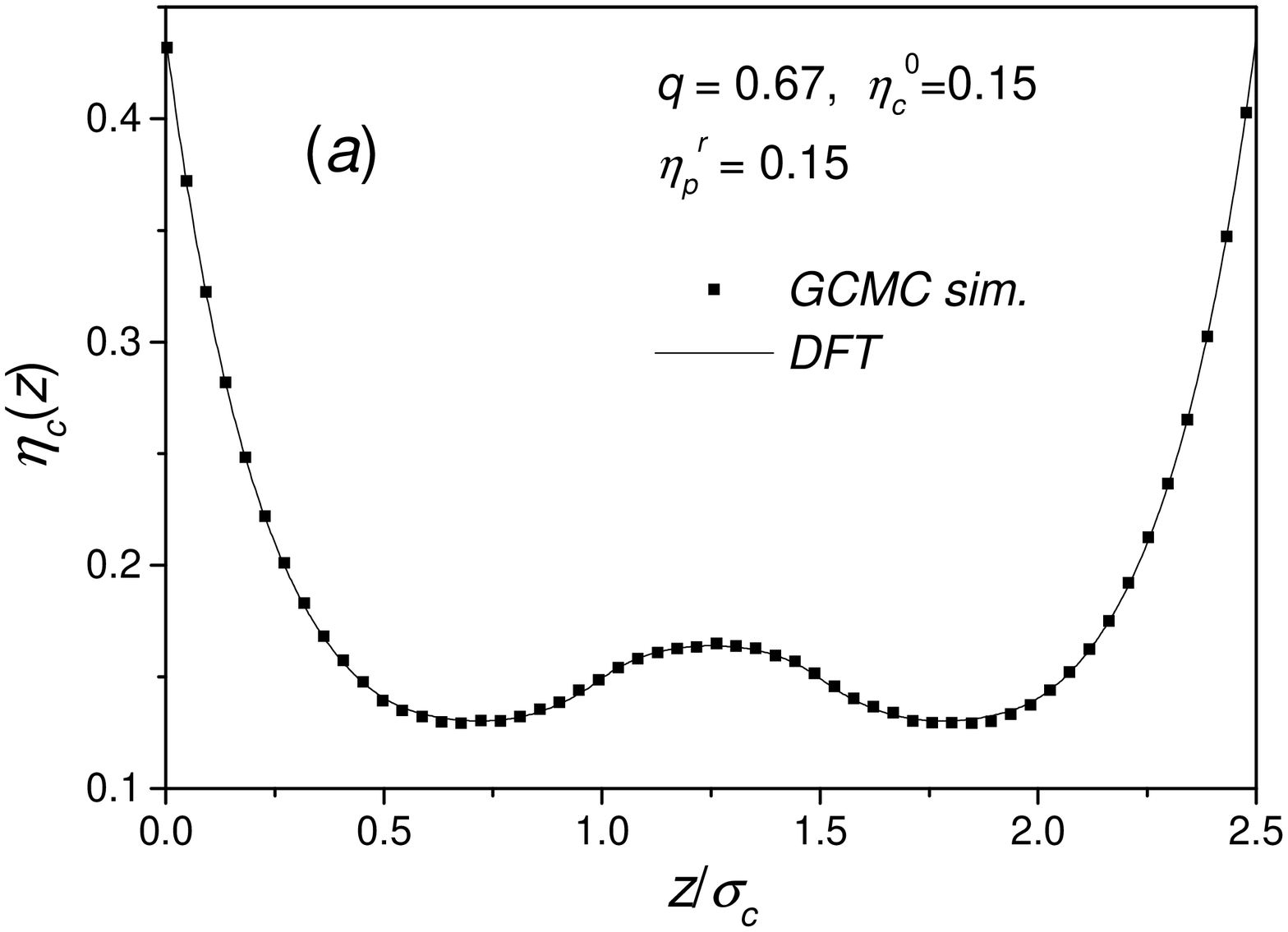}}
\center\resizebox{0.47\textwidth}{!}{\includegraphics{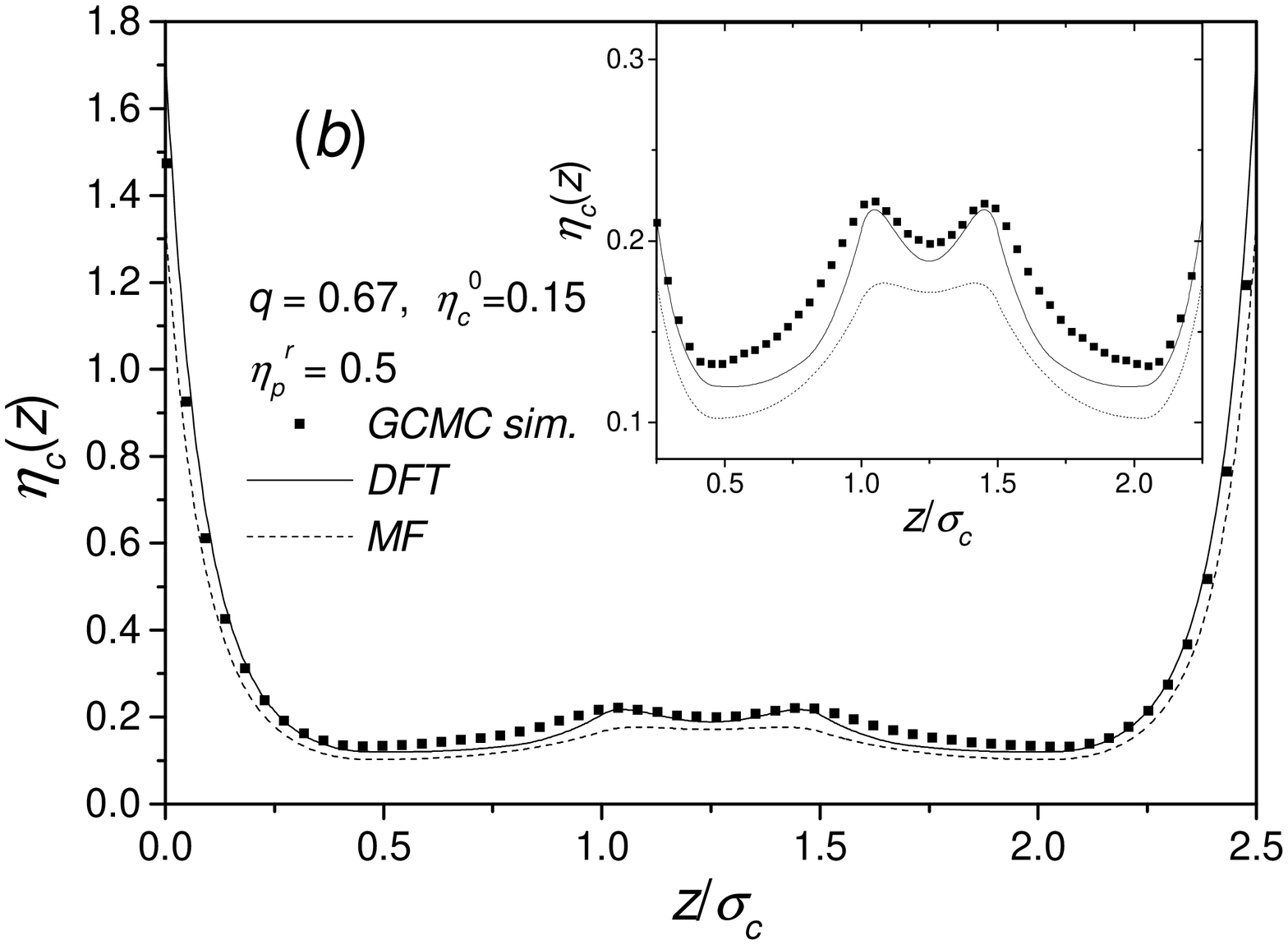}}  
\caption{\label{SIM5} (a) Colloid packing fraction between two hard walls obtained using GCMC simulations (squares) and DFT (solid line) for a mixture of colloids and interacting polymers and for $q=0.67$, $\eta_c^0=0.15$ and $\eta_p^r=0.15$. (b) Same as Figure~\ref{SIM5}a, but with $\eta_p^r=0.5$. Dashed lines are the results obtained from the mean--field DFT theory. The inset shows a magnification of the inner part of the data for better comparison.}
\end{figure}

The basic input in the evaluation of surface tensions or depletion forces are the density profiles $\rho_c(z)$. In order to check the reliability of the profiles calculated within the present perturbation DFT, we have carried out some grand canonical Monte--Carlo (GCMC) simulations of profiles, using the same effective colloid--colloid and colloid--wall interactions as with DFT. Two typical examples of such a comparison are shown in Figure~\ref{SIM5}a,b for a mixture of colloids and interacting polymers confined between two hard walls (slit geometry). At the lower polymer concentrations, corresponding to relatively weak colloid--colloid and wall--colloid attraction, the agreement between DFT and simulation is excellent (Figure~\ref{SIM5}a). At the higher polymer concentration (Figure~\ref{SIM5}b) the attractive perturbation is much stronger and, as expected, the agreement worsens. Nevertheless, the perturbation DFT is still capable of reproducing the main features of the density profile. Figure~\ref{SIM5}b also shows the result from the mean--field DFT, which amounts to neglecting correlations in $F_1\left[ \rho_c \right]$ (cf. section~\ref{wst}). The disagreement with the simulated data is now much more severe (see the inset in Figure~\ref{SIM5}b), stressing the importance of properly including reference system correlations in eq \ref{Psi1}.  

\section{The fluid--fluid interface}
\label{ff}

We finally turn to the fluid--fluid interface at coexistence between colloid--poor (\symbol{92}gas") and colloid--rich (\symbol{92}liquid") phases. This is a \symbol{92}free" interface, since translational symmetry is broken in the absence of any external potential, i.e.\ $\varphi_c(\vec{r})=0$. An accurate and consistent determination of the phase boundary (binodal) in the $\eta_c^0$ (colloid packing fraction)--$\eta_p^r$ (polymer reservoir packing fraction) is required. The bulk free energy per unit volume, obtained from eqs~\ref{FPDFT}--\ref{Psi1} in the homogeneous ($\rho_c(\vec{r})=\rho_c^0$) limit reads:
\begin{eqnarray}
\label{1orden}
\beta f &=& \rho_c^0\left[ \ln(\rho_c^0)-1 +\frac{\eta_c^0(4-3\eta_c^0)}{(1-\eta_c^0)^2}\right]  \\
&+& \frac{{(\rho_c^0)}^2}{2}\int d\vec{r}g_{HS}(r,\rho_c^0)\beta v(r).  \nonumber 
\end{eqnarray}
 This free energy depends implicitly on the polymer concentration
$\eta_p^r$ via the effective depletion pair potential
$v(r)=v(r;\eta_p^r)$, which is given by eq~\ref{vAO} ($v_{AO}$) for
ideal polymers and by eq~\ref{Vs} ($v_{int}$) for interacting
polymers.

For any fixed value of $\eta_p^r$ the colloid packing fractions of the
 coexisting phases may be determined by applying the standard Maxwell
 double--tangent construction to $f$. The resulting binodals for ideal
 and interacting polymers (not shown) are very close to those
 calculated in ref \cite{Rote04}, where an estimate of the second
 order perturbation correction to the free energy was included. The
 trends of the binodals are according to expectation. Qualitatively,
 the behaviour for ideal and interacting polymers is similar, with
 critical points at higher $\eta_p^r$ as $q$ increases, but with considerably
 \symbol{92}flatter" binodals for interacting polymers. There are, however, very significant
 quantitative differences, since the binodals for interacting polymers
 are shifted to considerably larger values of $\eta_p^r$ for each size
 ratio $q$, i.e.\ polymer interactions enhance the miscibility of
 colloid--polymer mixtures. This is easily understood, since the
 polymer--induced depletion attraction between colloids is weaker (for
 given $q$ and $\eta_p^r$) for interacting polymers.

\begin{figure} 
\center\resizebox{0.47\textwidth}{!}{\includegraphics{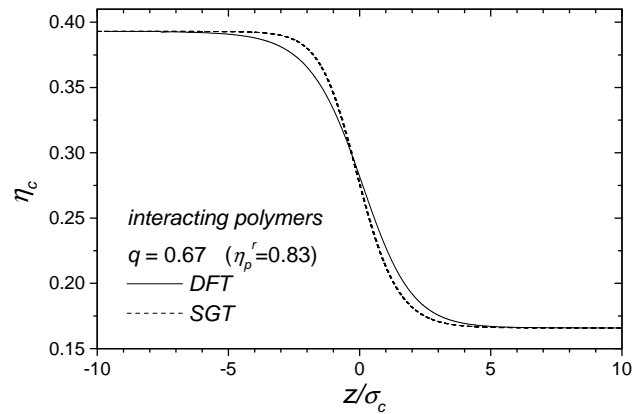}} 
\caption{\label{FF_LOU} Comparison between the colloid density profile for {\bf interacting} polymers obtained using our perturbative DFT (solid line) and the square gradient approximation SGT(dashed line), for $\eta_p^r=0.83$ and $q=0.67$ (close to the triple point).} 
\end{figure}

\begin{figure} 
\center\resizebox{0.47\textwidth}{!}{\includegraphics{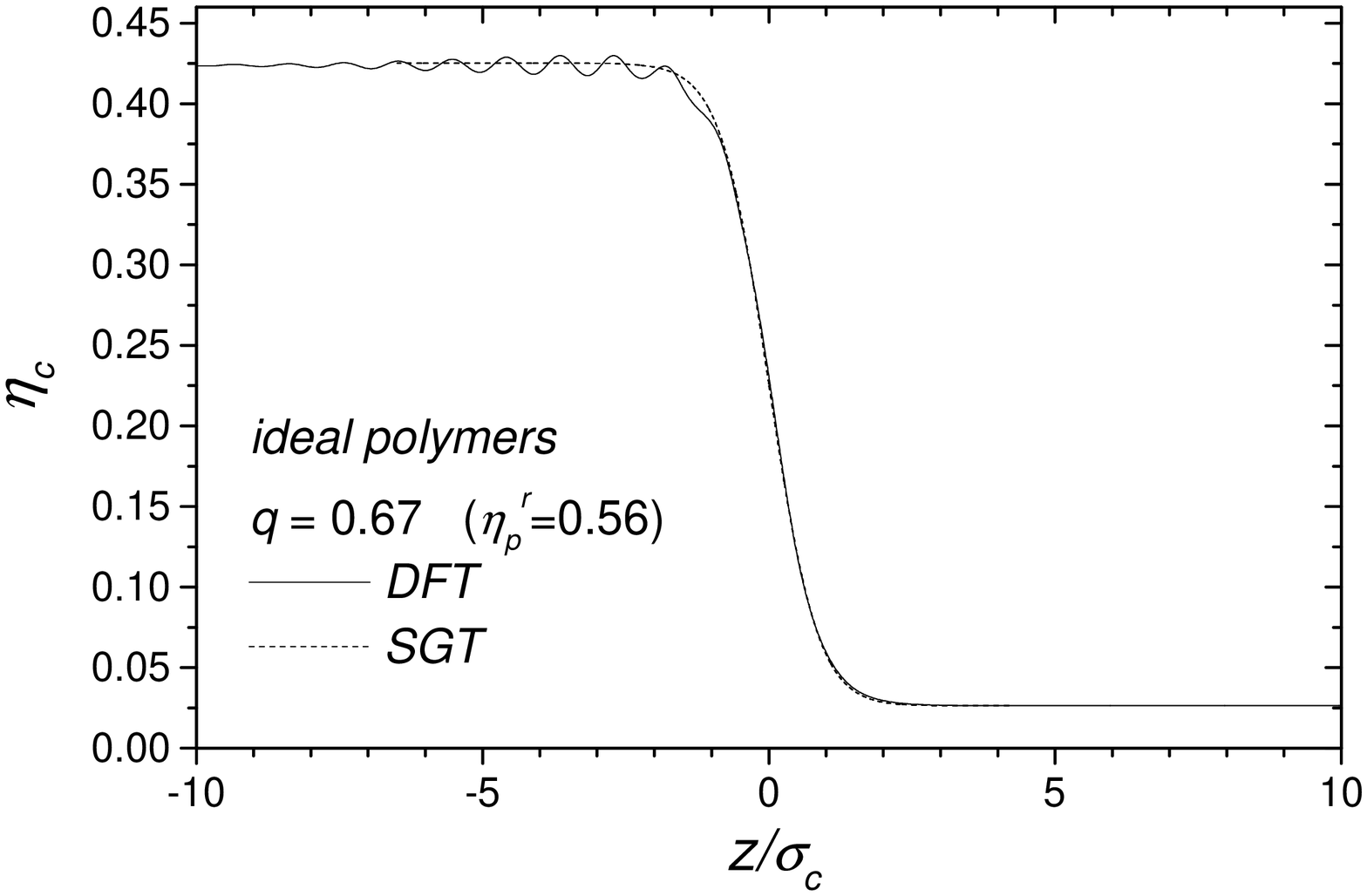}} 
\caption{\label{FF_AO} Comparison between the colloid density profile for {\bf ideal} polymers obtained using our perturbative DFT (solid line) and the square gradient approximation SGT (dashed line), for $\eta_p^r=0.56$ and $q=0.67$ (close to the triple point).} 
\end{figure}

The colloid density profile at the fluid--fluid interface is calculated by minimizing the grand potential (eq~\ref{omegagrand}), with $\varphi_c \equiv 0$, which leads back to the expression~\ref{rhorho0} (again with $\varphi_c \equiv 0$). The profiles for ideal and interacting polymers are shown in Figures~\ref{FF_LOU} and \ref{FF_AO}, under conditions close to the fluid--fluid--solid triple point, for a size ratio $q=0.67$. The profiles are compared to earlier predictions based on a square gradient density functional (SGT) \cite{Monc03}, using the same effective one--component description. There are remarkable differences between the profiles corresponding to interacting and ideal polymers. The interface is much sharper in the latter case and exhibits striking oscillations on the high colloid density (\symbol{92}liquid") side~\cite{Brad02}. The profile obtained with interacting polymers varies more smoothly, and shows no sign of oscillations. This clear--cut difference in behaviour may be partly understood by noting that the jump in colloid density between the two fluid phases is significantly smaller in the case of interacting polymers, for which the triple and critical points are much closer than for ideal polymers. Figures~\ref{FF_LOU} and \ref{FF_AO} also show the result of our earlier square gradient calculations (SGT) \cite{Monc03}. The corresponding profiles agree reasonably well with the results of the more elaborated density functional used in this paper; as expected the square gradient functional cannot account for the oscillatory structuring on the dense fluid side in Figure~\ref{FF_AO}. 


\begin{figure} 
\center\resizebox{0.47\textwidth}{!}{\includegraphics{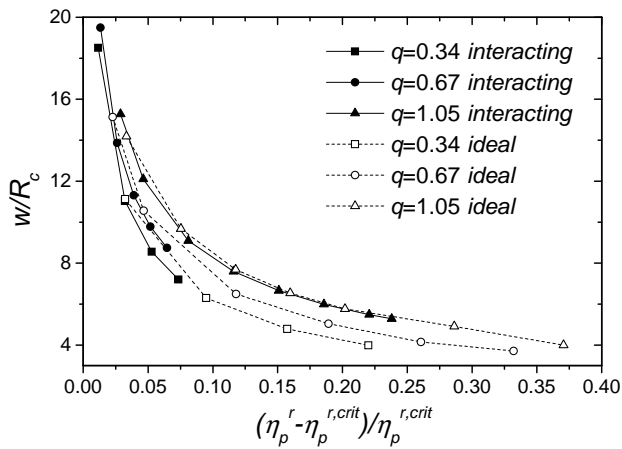}} 
\caption{\label{FF_W} Interfacial width obtained using the perturbation DFT for interacting polymers (solid lines and black symbols) and ideal polymers (dashed lines and white symbols), and for $q=0.34$, $0.67$ and $1.05$.} 
\end{figure}

In the case of interacting polymers (Figure~\ref{FF_LOU}) the width of the interface predicted by the present perturbation DFT is significantly larger than that obtained from square gradient theory. The interfacial width $w$ is conventionally defined as the distance between the two points where the density profile $\rho_c(z)$ reaches 90\% and 10\% of the difference between the bulk densities of the two coexisting phases. Results for $w$ obtained within the perturbation DFT are plotted in Figure~\ref{FF_W} as a function of the deviation of the polymer reservoir packing fraction from its value at the critical point, $\alpha = (\eta_p^r-\eta_p^{r,crit})/\eta_p^{r,crit}$, for ideal and interacting polymers and for three size ratios $q$. As expected, the width increases with $q$ and increases sharply as $\alpha$ decreases, i.e.\ upon approaching the critical point. Although the widths $w$ for ideal and interacting polymers are rather close for given $\alpha$ and $q$, polymer interactions tend to reduce the width compared to the ideal case, as found in \cite{Monc03}.

\begin{figure} 
\center\resizebox{0.47\textwidth}{!}{\includegraphics{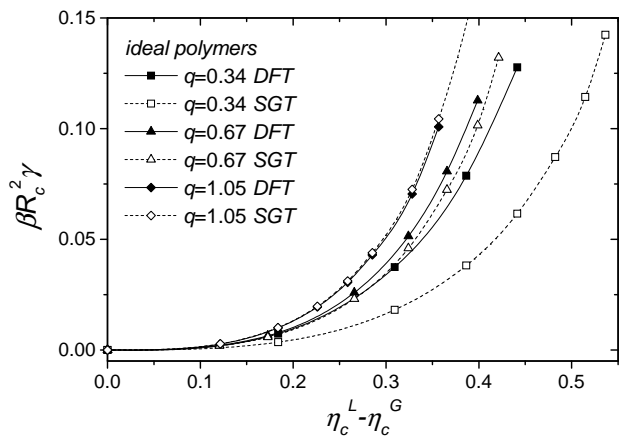}} 
\caption{\label{FF_GAMMA_AO} Comparison between the surface tension of the gas-liquid interface obtained from our perturbation DFT (solid lines and black symbols) and the square gradient theory SGT (dashed lines and white symbols), for ideal polymers.} 
\end{figure}

The surface tension $\gamma$ of the fluid--fluid interface is given by a relation similar to eq~\ref{gammaDFT}, and it is easily verified that the bulk terms $\Omega_0$ and $\Omega_1^{bulk}$ in eq~\ref{omegagrand} do not contribute to $\gamma$. The relation reads:
\begin{equation}
\gamma = \int_{-\infty}^{\infty}dz \{ f\left[ \rho_c(z) \right] -\rho_c(z)\mu_c^0 + P_c\}
\end{equation}
where $\mu_c^0$ and $P_c$ are the common values of the colloid contribution to the chemical potential and of the pressure in the coexisting bulk phases. Figure~\ref{FF_GAMMA_AO} compares results obtained from the perturbation and square gradient theories for ideal polymers, at three different size ratios; $\gamma$ is plotted versus the difference between the colloid packing fractions in the \symbol{92}liquid" and \symbol{92}gas" phases. The surface tensions are seen to increase with $q$ and $\eta_c^L-\eta_c^G$. The agreement between the two theories is excellent for $q=1.05$, and deteriorates for smaller size ratios. The surface tensions obtained from perturbation DFT are systematically larger than their square gradient counterparts and closer to the results of the two--component DFT of ref~\cite{Brad02}.

\begin{figure} 
\center\resizebox{0.47\textwidth}{!}{\includegraphics{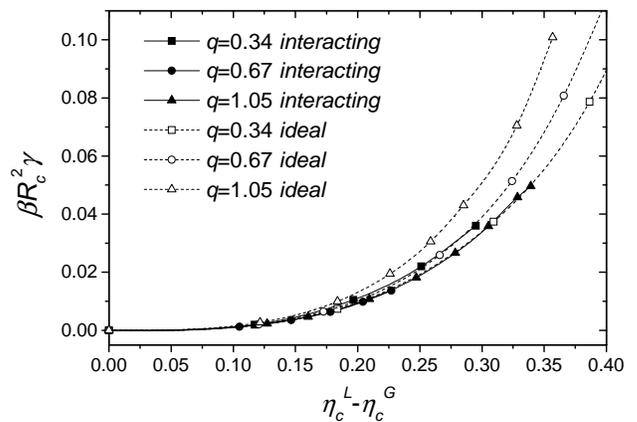}} 
\caption{\label{FF_GAMMA} Surface tension of the free liquid-gas interface for interacting polymers (solid lines and black symbols) and ideal polymers (dashed lines and white symbols), and for $q=0.34$, $0.67$ and $1.05$.} 
\end{figure}

A comparison between the surface tensions calculated within perturbation DFT for ideal and interacting polymers is finally made in Figure~\ref{FF_GAMMA}. The surface tension is seen to be lowered when polymer interactions are included, a trend which is opposite to that observed for the wall surface tension $\gamma_w$ (cf. Figure~\ref{GAMMAW10}).
A fluid-fluid interface also implies an inhomogeneous polymer density
profile.  For ideal polymers at constant $\mu_p$ this does not cost
any additional free-energy, but this is not true for interacting
polymers. For the latter depletant, this correction will raise the
surface tension compared to the values we calculate within the
one-component picture. On the other hand, for ideal polymers, the
many-body interactions between the colloids, fully taken into account
with the two-component formulation, also result in a larger surface
tension.  The effect of many-body interactions for interacting
polymers is unknown.  Thus our prediction for the relative strength of
the surface tensions should be tempered by the fact that the exact
form of the neglected corrections is not know and awaits a full
two-component treatment of interacting polymer-colloid mixtures.

\section{Conclusion}
\label{conc}

We have introduced a simple generic density functional description of
  colloid--polymer mixtures within the effective one--component
  picture where polymer degrees of freedom have been traced out. The
  free energy functional accounts correctly for the colloid excluded
  volume effects, and treats the polymer--induced depletion attraction
  within first order perturbation theory. The DFT formulation is very
  flexible and can be applied to confined colloid--polymer mixtures as
  well as to the free interface between coexisting fluid
  phases. Although this effective one--component description accounts
  only for pair--wise additive depletion interactions, and neglects
  more--than--two--body effective interactions (which are
  automatically included in an effective two--component
  representation), the present theory has the advantage of being able
  to treat the cases of ideal as well as interacting polymers
  consistently, thus allowing a direct estimate of the effect of
  polymer interactions on the interfacial properties of
  colloid--polymer mixtures. The present effective one--component DFT
  description can be tested against the more fundamental
  two--component DFT picture in the case of ideal
  polymers\cite{Brad02,Wess04}. Reasonable agreement is found for size
  ratios up to $q \approx 1$, giving confidence in the predictions of
  the simpler one--component representation for ideal and interacting
  polymers alike. The reduction to an effective one--component
  representation leads to the appearance of zero and one--body bulk
  contributions to the total grand potential of the mixture (terms
  $\Omega_0$ and $\Omega_1^{bulk}$ in eq~\ref{omegagrand}), which make
  significant contributions to the wall surface tension $\gamma_w$ and
  to the depletion interaction between two walls, and must not be
  overlooked. The key findings of the present investigation may be
  summarized as follows:

\begin{itemize}

\item Density profiles $\rho_c(z)$ near a hard wall deviate
considerably from pure hard sphere behaviour at low colloid packing
fractions. Significant differences between the behaviour observed for
ideal and interacting polymers may be traced back to the weaker
depletion attraction induced by the latter.

\item When the bulk terms $\Omega_0$
and $\Omega_1^{bulk}$ in the grand potential are properly taken into
account,  the wall surface tensions $\gamma_w$ calculated with ideal
polymers are in good agreement with a recent full two--component
treatment of the AO model\cite{Wess04}, for $q \leq 1$ and not too
high colloid packing fraction $\eta_c$. 

\item The depletion potential induced by colloid--polymer mixtures
between two walls exhibits structure on both length scales $R_c$ and
$R_p$ which opens the possibility of further flexibility in
\symbol{92}engineering" effective depletion forces.

\item In general, the wall surface tension $\gamma_w$ induced by
mixtures of colloids and interacting polymers is larger than that
found for ideal polymers. This trend is opposite to that predicted for
the fluid--fluid interfacial tension. The present predictions agree at
least qualitatively with the experimental data of ref~\cite{Aart03},
and may be understood in terms of the wall surface tension of the pure
polymer component.

\item The present effective one--component DFT predicts an oscillatory
density profile on the high colloid density side of the fluid--fluid
interface in the case of ideal polymers sufficiently close to the
triple point; this behaviour agrees with the earlier prediction based
on the effective two--component (AO model)
representation\cite{Brad02}. The oscillations are not observed in the
case of interacting polymers, presumably due to the smoother variation
(larger width) of the interfacial profile.

\item A direct comparison between the results of square gradient
analysis\cite{Monc03} and the present more sophisticated DFT shows
that the former is surprisingly accurate, even for rather sharp
fluid--fluid interfaces (except, of course, as regards the oscillatory
behaviour observed for ideal polymers). The good agreement validates
the predictions of the simple square gradient theory concerning the
effect of polymer--polymer interactions on the fluid--fluid
interfacial properties\cite{Monc03}.

\end{itemize}

The present effective one--component description predicts a number of
significant differences between interfacial behaviour of mixtures of
colloids and ideal versus interacting polymer coils. The corresponding
two--component representation involving ideal polymers (i.e.\ the AO
model) is well established, and confirms many of the predictions of
the present effective one--component picture. Future work should focus
on developing a viable effective two--component representation in the
case of interacting polymers, in order to validate the present
predictions based on the effective one--component description.

\acknowledgements

A. Moncho--Jord\'{a} thanks the Ministerio de Ciencia y Tecnolog\'{\i}a [Plan Nacional de Investigaci\'{o}n Cient\'{\i}fica, Desarrollo e Innovaci\'{o}n Tecnol\'{o}gica (I+D+I), project MAT 2003--08356--C04--01], J. Dzubiella acknowledges the EPSRC within the Portfolio grant RG37352, and A. A. Louis thanks the Royal Society (London), for financial support. Part of this work was carried out while J. P. Hansen was on leave at Universit\`{a} degli studi di Roma \symbol{92}La Sapienza", and support of INFM is gratefully acknowledged.

\section{Appendix}

\subsection{Fundamental Measure Theory for hard spheres.}
\label{appendixA}

In order to model the hard--sphere nature of the colloidal particles, we have used the White-Bear version of the Rosenfeld functional\cite{Roth02}. This functional improves the one proposed by Rosenfeld\cite{Rose89} in that it leads to the Mansoori et al.\ equation of state for a homogeneous mixture\cite{Hans86}. The free energy density $\Phi_{FMT} = \Phi_1+\Phi_2 +\Phi_3 $ reads\cite{Roth02}:
\begin{eqnarray}
\Phi_1 &=& -n_0\ln(1-n_3) \nonumber \\
\Phi_2 &=& \frac{n_1n_2-\vec{n}_{V1} \cdot \vec{n}_{V2}}{1-n_3} \\
\Phi_3 &=& \left( n_2^3-3n_2\vec{n}_{V2}^2\right)\frac{n_3-(1-n_3)^2\ln(1-n_3)}{36\pi n_3^2(1-n_3)^2} \nonumber
\end{eqnarray}
where $\{ n_j(\vec{r}) \}$ are weighted densities, obtained as convolutions of the colloid density $\rho_c(\vec{r})$ and the weight functions $\omega^{(j)}$ (see eq~\ref{nj}). The latter are given by:
\begin{eqnarray}
\omega^{(3)}(r) &=& \theta(R_c-r) \nonumber \\
\omega^{(2)}(r) &=& \delta(R_c-r) \nonumber \\
\omega^{(V2)}(\vec{r}) &=& (\vec{r}/r)\delta(R_c-r) \\
\omega^{(0)}(r) &=& \frac{\omega^{(2)}(r)}{4\pi R_c^2} , \ \omega^{(1)}(r)=\frac{\omega^{(2)}(r)}{4\pi R_c} , \ \omega^{(V1)}(\vec{r}) = \frac{\omega^{(V2)}(\vec{r})}{4\pi R_c} \nonumber
\end{eqnarray}

\subsection{Particular case: planar geometry.}
\label{appendixB}

For the case of planar fluid--fluid interfaces or infinite planar walls, the calculations simplify to a one-dimensional problem. Then, the position--dependent excess chemical potential $\mu_c^{ex}(\vec{r})$, which can be obtained from the functional derivative of the excess part of our perturbation free energy functional (eq~\ref{mucex}), depends only on $z$, the distance to the wall (or to the interface). Its explicit form for $z=z_1$ is\cite{Wade00}:
\begin{eqnarray}
\label{pdd}
\beta \mu_c^{ex}(z_1) &=& \int \sum_j \frac{\partial \Phi_{FMT}\left[ n_j(z_2) \right]}{\partial n_j(z_2)}\frac{\delta n_j(z_2)}{\delta \rho_c(z_1)}dz_2  \\
&+& 2\pi \int_{z_1-r_m}^{z_1-r_m}dz_2\rho_c(z_2) \times \nonumber \\
&\times& \int_{|z_1-z_2|}^{r_m}dr_{12}r_{12}\beta v(r_{12})g_{hs}(r_{12},\bar{\rho}_c(z_1,z_2))+ \nonumber \\
&+& \frac{3\pi}{4R_\nu^3}\int_{z_1-r_m}^{z_1-r_m}dz_2\rho_c(z_2)\left[ R_{\nu}^2-(z_1-z_2)^2 \right]\times \nonumber \\
&\times& \int_{z_2-r_m}^{z_2-r_m}dz_3\rho_c(z_3) \times \nonumber \\
&\times& \int_{|z_2-z_3|}^{r_m}dr_{23}r_{23}\beta v(r_{12})\frac{\partial g_{hs}(r_{23},\bar{\rho}_c(z_2,z_3))}{\bar{\rho}_c} \nonumber
\end{eqnarray}
where $v(r)$ is the corresponding colloid-colloid depletion potential, $r_m$ its range and
\begin{equation}
\frac{\delta n_j(z^{\ \prime})}{\delta \rho_c(z)}=\omega^{(j)}(z^{\ \prime}-z)
\end{equation}

For planar geometry, the weight functions are given by:
\begin{eqnarray}
\omega^{(0)}(z^{\ \prime}-z) &=&\frac{1}{2R_c}\theta(R_c-|z-z^{\ \prime}|) \nonumber \\
\omega^{(1)}(z^{\ \prime}-z) &=&\frac{1}{2}\theta(R_c-|z-z^{\ \prime}|) \nonumber \\
\omega^{(2)}(z^{\ \prime}-z) &=& 2\pi R_c\theta(R_c-|z-z^{\ \prime}|) \nonumber \\
\omega^{(3)}(z^{\ \prime}-z) &=& \pi\left[ R_c^2-(z-z^{\ \prime})^2\right]\theta(R_c-|z-z^{\ \prime}|) \nonumber \\
\vec{\omega}^{(V1)}(z^{\ \prime}-z) &=&\frac{1}{2R_c}(z-z^{\ \prime})\theta(R_c-|z-z^{\ \prime}|)\vec{k} \nonumber \\
\vec{\omega}^{(V2)}(z^{\ \prime}-z) &=&2\pi(z-z^{\ \prime})\theta(R_c-|z-z^{\ \prime}|)\vec{k} \nonumber
\end{eqnarray}

The $z$--dependent colloid density profile is obtained as
\begin{equation}
\label{rhorho0z}
\rho_c(z)=\rho_c^{0}\exp \{-\beta \left[\varphi_c(z)+\mu_c^{ex}(z))- \mu_c^{0,ex}  \right] \}
\end{equation}

The coupled equations eqs~\ref{pdd} and \ref{rhorho0z} are solved iteratively.

\end{document}